\documentclass[prd,twocolumn,showpacs,amsmath,amssymb]{revtex4}
\usepackage{amsmath}
\usepackage{amssymb}
\usepackage{mathrsfs}
\usepackage{bm}
\usepackage{txfonts}
\usepackage{graphicx}
\usepackage{booktabs}
\usepackage{dcolumn}
\usepackage{array}     
\usepackage{makecell}    
\begin{document}

\title{Dynamics of potential-free warm $\mathbf{k}$-inflation with nonminimal derivative coupling}
\author{Xiao-Min Zhang}
\thanks{Corresponding author}
\email{zhangxm@mail.bnu.edu.cn}
\affiliation{School of Science, Qingdao University of Technology, Qingdao 266520, China}
\author{Zi-Xin Bai}
\affiliation{School of Science, Qingdao University of Technology, Qingdao 266520, China}
\author{Run-Qing Zhao}
\affiliation{School of Science, Qingdao University of Technology, Qingdao 266520, China}
\author{Peng-Cheng Chu}
\affiliation{School of Science, Qingdao University of Technology, Qingdao 266520, China}
\author{Yun-Cai Feng}
\affiliation{School of Science, Qingdao University of Technology, Qingdao 266520, China}
\author{Zhi-Peng Peng}
\thanks{Corresponding author}
\email{zhipeng@mail.bnu.edu.cn}
\affiliation{School of Physics and Advanced Energy, Henan University of Technology, Zhengzhou 450001, China}
\author{Xi-Bin Li}
\thanks{Corresponding author}
\email{lxbimnu@imnu.bnu.edu.cn}
\affiliation{College of Physics and Electronic Information, Inner Mongolia Normal University, 81 Zhaowuda Road, Hohhot, 010022, Inner Mongolia, China}
\date{\today}

\begin{abstract}
In contrast to potential-driven warm inflation models, this paper presents a new inflationary scenario driven purely by noncanonical kinetic terms. We derive the evolution equations and the associated slow-roll approximations specific to the kinetic case. The model incorporates a nonminimal derivative coupling that enhances gravitational friction; when combined with thermal damping, this leads to a significantly slower evolution of the pure kinetic inflaton. The resulting slow-roll approximations differ fundamentally from those of potential-driven inflation. The attractor behavior of this warm $k$-inflation with nonminimal derivative coupling is explored, confirming that slow-roll solutions can approach a strict exponential expansion attractor under relaxed slow-roll conditions. We further calculate the density fluctuation equations and obtain analytic expressions for the power spectrum, spectral index, and tensor-to-scalar ratio. Compared to standard inflation in general relativity, the energy scale at horizon crossing is lower, and the tensor-to-scalar ratio is significantly reduced due to the combined effects of thermal damping and nonminimal derivative coupling. The field excursion remains comfortably sub-Planckian. The model's predictions are in excellent agreement with the latest Planck 2018 data, offering a novel and successful extension of the warm inflation paradigm.
\end{abstract}
\maketitle

\section{\label{sec:level1}Introduction}

Inflation, a brief epoch of quasi-exponential expansion in the early universe, effectively addresses the horizon, flatness, and monopole problems \cite{Guth1981,Linde1982,Albrecht1982,Bassett2006}. During this accelerated phase, quantum fluctuations naturally gave rise to the temperature anisotropies and large-scale structures observed today \cite{PLANCK1,PLANCK2}. In addition to the traditional cold (standard) inflationary models, warm inflation has emerged as a strong alternative for cosmic inflation theory. Introduced three decades ago \cite{BereraFang,Berera1995}, warm inflation involves continuous energy exchange between the inflaton and a sub-dominant bath of bosonic or fermionic degrees of freedom, governed by an interaction Lagrangian $\mathcal{L}_{\text{int}}$, maintaining a non-zero temperature throughout the expansion. Unlike standard inflation, which requires a distinct reheating phase, warm inflation allows for a seamless transition to radiation domination. Notably, thermal fluctuations, rather than vacuum fluctuations, dominate the primordial spectrum \cite{BereraFang,Lisa2004,Berera2000,Moss2009,WarmSPy2024}, and the associated dissipative damping alleviates stringent flatness conditions on the potential, addressing both the $\eta$ problem \cite{etaproblem,BereraIanRamos} and the excessive super-Planckian field excursions typically found in cold inflation \cite{Berera2005,BereraIanRamos}.

To date, inflation has predominantly been developed within the framework of Einstein's general relativity (GR), with few attempts to extend it to modified gravity theories. One significant category of modified gravity is nonminimal coupling theory, which offers two main approaches. The first approach involves a generic function $f(\phi)$ that couples to the Ricci scalar, enabling inflation to directly source curvature \cite{Kaiser1995, Karydas2102.08450, EadkhongNPB2023}. The second approach, followed in this paper, couples the inflaton kinetic term to the gravitational sector via a nonminimal derivative coupling (NMDC) to the Einstein tensor \cite{Karydas2102.08450,GermaniPRL2010,DalianisJCAP2020,Sadjadi2015,Han2025}. The NMDC term, $G^{\mu\nu}\partial_{\mu}\phi\partial_{\nu}\phi$ \cite{GermaniPRL2010,YangNan2015,Sadjadi2015,Zhang2025}, generates enhanced gravitational friction that decelerates the scalar field's evolution. Initially proposed for cold inflation \cite{GermaniPRL2010, DalianisJCAP2020, Sadjadi2015}, this coupling has now been extended to the warm inflationary regime, leading to a potential-dominated noncanonical warm-NMDC scenario in previous studies \cite{Zhang2025}. The resulting field equations remain second-order, and the coupling introduces gravitational friction that complements, and often dominates, the usual Hubble damping. Importantly, once the kinetic term of the scalar field is coupled nonminimally with the Einstein tensor, the effective Higgs self-coupling $\lambda$ drops to order unity, avoiding the introduction of higher degrees of freedom \cite{GermaniPRL2010,YangNan2015}. If the scalar field is massless and does not exhibit the canonical kinetic term $g^{\mu\nu}\partial_{\mu}\phi\partial_{\nu}\phi$, its derivative coupling to gravity can resemble the behavior of dark matter \cite{Gao2010,Ghalee2013}.

Previous studies have shown that the introduction of the thermal dissipative term $\Gamma\dot\phi$, characteristic of warm inflation, revives the quartic potential $\lambda\phi^{4}$, which has long been excluded in general relativity cold inflation. This makes the model consistent with current observational data \cite{YangNan2015,BeingWarm2014}. A thorough examination shows that the combined gravitational and thermal friction enforces the slow-roll conditions largely independent of the potential's detailed form, tightens the Lyth bound, and ensures that the field excursion stays well below the Planckian scale, ensuring both theoretical consistency and observational viability \cite{Zhang2025}.

The concept of $k$-inflation, first proposed in \cite{Mukhanov1999,Mukhanov1999-1}, revolutionized the inflationary paradigm by illustrating that accelerated expansion can be driven entirely by noncanonical kinetic terms, eliminating the need for a flat potential. Rooted in string-theoretic constructions such as Dirac-Born-Infeld dynamics, $k$-inflation assigns the inflaton a Lagrangian density $\mathcal{L}(\phi, X)$, where $X = \frac{1}{2}g^{\mu\nu}\partial_\mu\phi\partial_\nu\phi$, and a sound speed $c_s^2 = p_{,X}/\rho_{,X}$ that can significantly differ from unity. The $k$-inflation model can still exhibit ``slow-roll" behavior, where the slow-roll conditions need not be strictly fulfilled \cite{Mukhanov1999,Peng2016}. Moreover, this extra degree of freedom alters the amplitude and shape of primordial fluctuations, leading to modified scalar-tensor consistency relations and creating possibilities for large, observationally accessible non-Gaussianities \cite{Mukhanov1999-1,Peng2018,Grigoris2007,Arroja2011,Zhang2023}.

In this paper, we extend the $k$-inflation paradigm to the warm inflationary framework. By replacing the super-cooling and abrupt reheating characteristic of cold inflation with a thermal bath maintained by dissipative inflation interactions, the model leads to different thermal perturbations \cite{Peng2016,Peng2018}. The presence of NMDC, which enhances gravitational damping, and the inclusion of thermal dissipation terms makes it natural to investigate the consequences of extending purely kinetic-driven inflation to the warm inflation framework with NMDC. While previous research has addressed NMDC warm inflation driven by a potential \cite{Zhang2025}, this study focuses on the distinctive pure kinetic regime where inflation is driven solely by noncanonical kinetic terms, with no potential term. A critical question arises: how do these two damping mechanisms interact and jointly affect field evolution, slow-roll approximations, and the primordial perturbation observables? The following sections will explore the application of this distinctive purely kinetic scalar field within the warm inflation framework with nonminimal derivative coupling. We systematically analyze warm $k$-inflation scenarios augmented by NMDC, and the resulting framework generates predictions that are consistent with current observations, while presenting several clear advantages over traditional inflationary models. As will be shown, the model naturally accommodates the Planck 2018 constraints on the spectral index and the tensor-to-scalar ratio across a broad and physically well-motivated region of parameter space.

The paper is organized as follows.
In Sec. \ref{sec:level2}, we present the general framework of NMDC warm $k$-inflation and derive the background dynamical equations. Sec. \ref{sec:level3} is devoted to establishing the slow-roll conditions and analyzing the stability of the attractor solution. In Sec. \ref{sec:level4}, we investigate the cosmological perturbations and obtain the resulting scalar spectrum, spectral index, and related quantities. Sec. \ref{sec:level5} provides a concrete realization of the model and confronts its predictions with current observational constraints from Planck data. Finally, Sec. \ref{sec:level6} summarizes the main results and discusses their physical implications and possible future directions.

\section{\label{sec:level2} NMDC Warm $\mathbf{K}$ Inflationary Scenario}

Now, this paper proposes the framework of this new NMDC warm $k$-inflationary model.
The model is based on a distinctive nonminimally derivative-coupled Higgs theory linked to gravity \cite{GermaniPRL2010}, which does not introduce higher degrees of freedom than those in general minimal interaction with a scalar field. Considering the energy transfer from the inflation to radiation, the overall action of the multi-component NMDC warm inflationary case is:
\begin{equation}\label{action}
    S=\int d^4x\sqrt{-g}\left[\frac{1}{2}M_{pl}^2R+\mathcal{L}(X,\phi)+\frac{G^{\mu\nu}}{2M^2}\partial_
    {\mu}\phi\partial_{\nu}\phi +\mathcal{L}_{int}+\mathcal{L}_R\right],
\end{equation}
where $G_{\mu\nu}$ denotes the Einstein tensor, $M_{pl}^2 \equiv {\left( {8\pi G} \right)^{ - 1}}$ presents the decreased squared Planck mass, and $M$ is the mass-dimension coupling constant. The total Lagrangian density can be represented as $\mathcal{L}_{total}=\mathcal{L}_g+\mathcal{L}(X,\phi)+\mathcal{L}_{NMDC}+\mathcal{L}_R+\mathcal{L}_{int}$, where $\mathcal{L}_g=\frac{1}{2}M_{pl}^2 R$ signifies the gravitational Lagrangian density, while $\mathcal{L}_{NMDC}=\frac1{2M^2}G^{\mu\nu} \partial_{\mu}\phi\partial_{\nu}\phi$ denotes the NMDC Lagrangian density. The Lagrangian density for the inflaton field is expressed as $\mathcal{L}(X,\phi)$, a general function of the field $\phi$ and the kinetic term $X$, and $\mathcal{L}(X,\phi)$ is often simplified to $\mathcal{L}$ for convenience. $\mathcal{L}_R$ represents the radiation fields, and $\mathcal{L}_{int}$ describes the interactions between the inflaton and the remaining subdominant components. In a traditional scalar field model with a potential, the Lagrangian density is expected to revert to the canonical form (i.e., $\mathcal{L}= X-V$) in the limit of small $X$, thus ensuring a consistent normalization of the field \cite{Franche2010}.

The equation of motion for the inflaton field dictates the primary evolution of the multi-component universal inflation, and it can be derived by varying the total action Eq. (\ref{action}):
\begin{equation}\label{EOM0}
\frac{\partial\left[\mathcal{L}(X,\phi)+\mathcal{L}_{int}\right]}{\partial\phi} -\left(\frac{1}{\sqrt{-g}}
\right)\partial_{\mu}\left[\sqrt{-g}\frac{\partial\left(\mathcal{L}(X,\phi)+\mathcal{L}_{NMDC}\right)}
{\partial(\partial_{\mu}\phi)}\right]=0.
\end{equation}

In the context of a homogeneous spatially flat Friedmann-Robertson-Walker (FRW) universe, the assumption of a homogeneous inflation mean field, i.e., $\phi=\phi(t)$ reduces the motion equation to

\begin{equation}\label{EOM1}
\begin{aligned}
    \left[\left(\frac{\partial\mathcal{L}(X,\phi)}{\partial X}\right)+2X\left(\frac{\partial^2\mathcal{L}(X,\phi)}{\partial X^2}\right)+\frac{3H^2}{M^2}\right]\ddot{\phi}\quad\quad\quad \\   +\left[3H\left(\frac{\partial\mathcal{L}(X,\phi)}{\partial X}+\frac{3H^2}{M^2}+\frac{2\dot H}{M^2}\right)\right]\dot{\phi}+\frac{\partial^2\mathcal{L}(X,\phi)}{\partial X\partial\phi}\dot\phi^2 \\ -\frac{\partial(\mathcal{L}(X,\phi)+
    \mathcal{L}_{int})}{\partial\phi}=0,\quad\quad\quad\quad\quad\quad\quad\quad\quad
\end{aligned}
\end{equation}
where $X=\frac12\dot\phi^2$ in the FRW universe.
By varying the universal action with respect to the metric, the field equations integrating the NMDC effect can be obtained \cite{Zhang2025}.

Having laid out the general tensor formalism for NMDC warm inflation in \cite{Zhang2025}, this study now focuses on its distinctive potential-free Lagrangian. Its Lagrangian density $\mathcal{L}(X,\phi)$ consists of only a kinetic term and is devoid of a potential term, depending functionally on $X$ and possibly on $\phi$ \cite{Mukhanov1999}. Given that a pure kinetic field serves effectively as an inflaton \cite{Mukhanov1999,Mukhanov1999-1,Peng2016}, it can be embedded into the NMDC warm inflation framework. The Lagrangian density for an original kinetic inflaton field can be expressed simply as in Ref. \cite{Mukhanov1999,Peng2016}: $p(\varphi ,X')=K(\varphi)X'+L(\varphi )X'^{2}+\cdots$ (the original pure kinetic field is represented as $\varphi$, and its kinetic term is denoted by $X'=\frac12\dot\varphi^2$). Here, the analysis is simplified by redefining the scalar field and using the new field variable $\varphi^{\rm new}=\int d\varphi^{\rm old}L^{1/4}(\varphi^{\rm old})$ (which is well defined under the assumption $L(\varphi)>0$). With this definition, we have $L_{\rm new}(\varphi^{\rm new})=1$
and $K_{\rm new}(\varphi^{\rm new})=K_{\rm old}(\varphi^{\rm old})/L_{\rm old}^
{1/2}(\varphi^{\rm old})$. Thus, without loss of generality, it can be assumed that $L(\varphi)=1$. Then, the Lagrangian density can be rewritten as: $\mathcal{L}_{new}=K_{new}(\varphi_{new})X_{new}+X^2_{new}$. For simplicity, the subscript ``new'' is omitted, and the newly defined inflaton is represented as $\phi$ hereinafter. To keep the inflaton field have a canonical dimension in mass, i.e., $[\phi]=[m]$, a constant $\mu$ with mass dimension can be introduced in the denominator of the second term as
$\mathcal{L}=K(\phi)X+X^2/\mu^4$. In this new expression, the function $K(\phi)$ is defined to incorporate a factor of $1/\mu^{2}$ compared to its original definition; for simplicity, the same notation is retained.
The Lagrangian density is then formulated as
\begin{equation}\label{Lphi}
\mathcal{L}(X,\phi)=K(\phi)X+X^2/\mu^4,
\end{equation}
where, $K(\phi)$, called ``kinetic function'', depends on the inflaton field, and its rate of change over a Hubble time is $\frac{\dot K}{HK}=\frac{K_{\phi}\dot\phi}{HK}$. For $X$ to remain positive in slow roll inflation, $K(\phi)$ must be negative, and a new parameter $\overline{K}(\phi)=-K(\phi)$ is defined, which is positive. The pressure of the inflaton is given by $p_{\phi}=\mathcal{L}(X,\phi)=K(\phi)X+X^2/\mu^4$, and the energy density can be expressed as $\rho_{\phi}=K(\phi) X+3X^2/\mu^4$. In the definitions of $p_{\phi}$ and $\rho_{\phi}$, the subscript $\phi$ is a label for identifying the inflaton field and does not indicate a derivative concerning $\phi$. A crucial parameter for the general noncanonical field, including the pure kinetic field, is the speed of sound, which describes the propagation speed of scalar perturbations: $c_{s}^{2}=p_{X}(\phi,X)/\rho_{X}(\phi,X)=\left(1+2X\mathcal{L}_{XX}/\mathcal{L}_{X}\right)^{-1}$, with $X$ referring to a derivative.

In the context of warm inflation, there exists an interaction term $\mathcal{L}_{int}$, as evidenced by Eqs. (\ref{action}), (\ref{EOM0}) and (\ref{EOM1}). The interaction term relies on the zeroth order of the inflaton and other fields, excluding their derivatives. Mechanisms that effectively describe the interaction between the inflaton and other fields comprise the supersymmetric two-stage model \cite{BereraKephart,MossXiong} and the warm little inflaton model \cite{Berera2016}. To accurately depict the dissipation of $\phi$ into other fields \cite{BereraFang,Berera2005,Berera2000,BereraIanRamos}, a term $\Gamma\dot\phi$ can be incorporated into the evolution equation similar to the GR frame warm inflation \cite{Zhang2014,BereraFang,Berera2016}, acting as a thermal damping term. The temperature associated with this term corresponds to that of the radiation bath, which is constantly nonzero, thereby keeping the inflationary universe warm owing to the inflaton's dissipation into the bath, given that the temperature dependence of the dissipation coefficient lies within the range $(T^{-4}\sim T^4)$ \cite{Ian2008,Campo2010,Zhang2014,Zhang2025}.

Under the assumptions of warm inflation and through variation calculations, the equation of motion is:
\begin{equation}\label{EOM2}
    (\mathcal{L}_{X}c_{s}^{-2}+3F)\ddot{\phi}+3H\left(\mathcal{L}_{X}+3F+\frac{2\dot{F}}{H}\right)\dot{\phi}
    +\Gamma\dot{\phi}+\frac12K_{\phi}\dot\phi^2=0,
\end{equation}
where $F=\frac{H^{2}}{M^{2}}$, and the term $\dot F\ll HF$ during inflationary epoch. The parameter $r=\Gamma/3H$ is traditionally used to describe the thermal dissipative strength in GR-based warm inflation, with $r\gg1$ and $r\ll1$ indicating strong and weak dissipative warm inflation, respectively. Compared to the preceding term, the term $\frac{2\dot{F}}{H}\dot{\phi}$ can be omitted, which can be assured by the slow roll condition of $\epsilon$ as discussed in the following section. When $F\ll1$, the Einstein GR limit is recovered, and $F\gg1$ denotes a regime of high gravitational friction limit. From the evolution equation above, the Hubble damping term is a factor of $\mathcal{L}_X$ larger than in canonical inflation. This, combined with the additional gravitationally enhanced and thermal friction, causes the inflaton field to evolve more slowly. Thus, the damping term is amplified by a factor of ($\mathcal{L}_{X}+3F+r$) in comparison to GR frame standard inflationary picture, suggesting that the NMDC warm inflationary case exhibits a serious over-damping phenomenon.

By integrating the Hamiltonian constraint with the warm universal radiation contribution, one can derive the Friedmann equation for the NMDC warm $k$-inflation as:
\begin{equation}
    3H^{2}=\frac{1}{M_{p}^{2}}(2X \mathcal{L}_{X}-\mathcal{L}+9FX+\rho_r).
\end{equation}
Taking into account all components of the warm universe, the total energy density can be explicitly represented as $\rho=\rho_{\phi}+9FX+\rho_r=KX+\frac{3\mathrm{X}^2}{\mu^4}+9FX+\frac{3}{2}rX$. Additionally, the pressure of the NMDC warm universe can be determined as $p=p_{\phi}-3FX-\frac{2\dot H}{M^2}-\frac{2H}{M^2}\dot\phi\ddot\phi+p_r\simeq KX+\frac{X^2}{\mu^4}-3FX+\frac{1}{2}rX$, using the warm slow roll approximation equation (\ref{SRentropy}) outlined in the subsequent section.

By integrating the inflaton evolution equation, along side the total energy conservation equation $\dot\rho+3H(\rho+p)=0$ and the radiation energy relation $\rho_r=3Ts/4$ of warm inflation, one can derive the entropy production equation:
\begin{equation}\label{Ts}
    T\dot{s}+3HTs=\Gamma\dot{\phi}^{2},
\end{equation}
where $s$ denotes the entropy density of the warm universe. Given that $\rho_r=3Ts/4$, Eq. (\ref{Ts}) transforms into an equation that governs the production of radiation energy density
\begin{equation}\label{rhor}
  \dot{\rho}_r+4H\rho _r=\Gamma \dot{\phi}^2.
\end{equation}

Inflation theory is inherently predictive, so it is typically related to slow roll approximations that yield quasi-exponential expansion. The slow roll regime in this purely kinetic-term-dominated inflationary model is the analog of ``slow roll" in potential-driven inflation, but it essentially differs from the latter. The zeroth-order slow-roll solution to the evolution equation (\ref{EOM2}), representing the dynamical system's instantaneous attractive fixed point, can be derived from the total energy conservation condition, $\rho + p = 0$. This condition indicates that $\rho$ decreases above the line $p = -\rho$ and increases below it. Consequently, the inflationary attractor is described by the following equations:
\begin{equation}\label{Xf}
  X_{f}=-\mu^4\frac{K(\phi_f)+3F+r}{2}=\mu^4\left(\frac{\overline{K}(\phi_f)}{2}-\frac{3F+r}{2}\right),
\end{equation}
\begin{equation}\label{dotphif}
  \dot\phi_f=\sigma\mu^2\sqrt{\overline{K}(\phi_f)-3F-r},
\end{equation}
where $\sigma$ denotes the sign of $\dot\phi$, and
\begin{equation}\label{srrho}
  \rho_f=\frac{X_f}{2}(9F+\overline{K}),
\end{equation}
\begin{equation}\label{Hf}
  H_f^2=\frac{1}{3M_p^2}\rho_f=\frac{1}{6M_p^2}X_f(9F+\overline{K}),
\end{equation}
where the subscript ``f'' signifies the quantities at the attractive fixed point. The actual slow roll solution does not need to match the attractor solution $X_f$, but should highly approach it: $X_{sr}\rightarrow X_f$, so they can be identified as one quantity (the subscript ``sr'' indicates the slow roll solutions). During inflationary epoch, $X=X_f+\delta X$, and the condition for the validity of our slow roll solution is given by:
\begin{equation}\label{sr}
  \frac{\delta X}{X_f}\ll1.
\end{equation}

With the new slow roll conditions, one can derive the time evolution and the number of e-folds of the slow roll NMDC warm $k$-inflation from the above equations:
\begin{equation}\label{t}
t-t_{\ast}=\sigma\int^{\phi}_{\phi_{\ast}}\frac{d\phi}{\sqrt{\overline{K}-3F-r}},
\end{equation}
\begin{equation}
N=\int H dt=\int\frac{H}{\dot{\phi}}d\phi\simeq\frac{\sigma}{\sqrt3M_p}\int_{\phi_{\ast}}^{\phi}\sqrt{9F+\overline{K}}d\phi,
\end{equation}
where $\phi_{\ast}$ signifies the field value at the Hubble horizon crossing, and $\phi_e$ represents the
final value of the inflaton field. The slow roll conditions achievable in this model, which will be thoroughly analyzed in the next section, ensure an adequate number of e-folds to tackle the horizon and flatness issues encountered in the standard model of cosmology.

\section{\label{sec:level3}Slow Roll and Consistency Analysis}
Inflation is sustained not by potential flatness, but by a dynamical balance among noncanonical kinetic structure, NMDC-enhanced gravitational friction, and thermal dissipation.

\subsection{Slow Roll Approximations and Conditions}

To ensure that the inflationary system remains in the slow-roll solution for a sufficiently long period, a detailed analysis is required. Here, we present the slow-roll approximations and conditions that ensure the system remains in the slow-roll regime. The following outlines the key slow-roll parameters pertinent to the NMDC warm $k$-inflationary model:
\begin{equation}
    \epsilon = -\frac{\dot{H}}{H^2}, \quad \eta = \frac{K_{\phi} u}{H K}, \quad \beta = \frac{\Gamma_{\phi} u}{H \Gamma},
\end{equation}
where $ u = \dot{\phi} $ is the inflaton velocity, and $ K_{\phi} $ and $ \Gamma_{\phi} $ are the relevant derivatives of the kinetic function and dissipation coefficient, respectively.

Additionally, we introduce a characteristic parameter related to temperature dependence:
\begin{equation}
    c = \frac{T \Gamma_T}{\Gamma},
\end{equation}
which quantifies the temperature dependence of the dissipation rate.

Using the new variable $ u $, where $ X = \frac{1}{2} u^2 $, the evolution equations governing the warm inflationary dynamics can be rewritten as:
\begin{equation}\label{dotu}
    (6X/\mu^4 + K) \dot{u} + 3H (2X/\mu^4 + K + 3F + r) u + K_{\phi} X = 0,
\end{equation}
and the entropy production equation becomes:
\begin{equation}\label{dots}
    \dot{s} = -3H s + \frac{\Gamma u^2}{T}.
\end{equation}

In the slow-roll regime, the evolution of the inflaton kinetic term, energy density, and Hubble parameter is expected to be gradual, with quasi-static radiation production. However, the slow-roll approximations in this model differ significantly from potential-dominated inflationary models, where the slow-roll condition typically involves neglecting higher-order terms in the evolution equations. In contrast, our model does not rely on truncating equations but rather involves a dynamical balance between contributions from the noncanonical kinetic term, the nonminimal derivative coupling, and the dissipation effects.

It is important to clarify the role of the slow-roll approximation in the present framework.
Unlike conventional potential-driven inflation, where slow-roll typically implies neglecting higher-order terms in the equation of motion, the situation in the present purely kinetic NMDC warm inflation model is fundamentally different.
In particular, near the attractor solution, the combination $(2X/\mu^4 + K + 3F + r) \sim \mathcal{O}(\delta X)$ is itself of perturbative order. As a consequence, the second term in Eq. (\ref{dotu}) is suppressed and becomes comparable to the first term $(6X/\mu^4+K)\dot{u}$. Therefore, the two terms are of the same order and must be retained simultaneously in a consistent linear perturbation analysis. This explains why, despite adopting a slow-roll regime characterized by $\dot{u} \ll H u$, it is not self-consistent to neglect the $\dot{u}$ term in Eq. (\ref{dotu}). The slow-roll condition in this model should instead be understood as a statement about the overdamped evolution of the system, rather than a truncation scheme of the dynamical equation.

This feature originates from the purely kinetic structure of the model and the existence of a kinetic attractor, and thus differs qualitatively from the standard slow-roll approximation in potential-driven inflation.

To clarify, the slow-roll solutions for this model are governed by the following conditions:
\begin{equation}\label{SREOM}
    \delta X \ll X_f,
\end{equation}
and
\begin{equation}\label{SRentropy}
    3H T s = \Gamma u^2.
\end{equation}

These conditions ensure that the system evolves very slowly and remains at the attractor point, with key conditions such as $ \dot{H}/H^2 \ll 1 $, $ \dot{\rho}/H\rho \ll 1 $, and $ \dot{s}/Hs \ll 1 $. In particular, these conditions imply that $ \dot{\rho}_r / 4H \rho_r \ll 1 $ and $ \dot{T}/HT \ll 1 $, which will be proved in the following sections.

From the energy conservation equation, we derive the following relation for $ \delta X $:
\begin{equation}\label{deltaX}
    \frac{\delta X}{X_f} = -\frac{\mu^4 \dot{\rho}}{12 H X_f^2} = -\frac{\dot{\rho}_f}{12H \rho_f} \cdot \frac{9F + \overline{K}}{\overline{K} - 3F - r}.
\end{equation}
Using the Friedmann equation $ \rho \propto H^2 $, we find that $ \dot{H}/H^2 \sim \dot{\rho}/H\rho $, and both are proportional to $ \delta X / X_f $.

By considering the slow-roll solution for $ X_f $, we derive the following relation:
\begin{equation}\label{deltaX1}
    \frac{\delta X}{X_f} = \frac{(\overline{K} - 3F - r)^{\bullet} (9F + \overline{K})}{12 H (\overline{K} - 3F - r)^2} + \frac{(9F + \overline{K})^{\bullet}}{12 H (\overline{K} - 3F - r)}.
\end{equation}
The dot here denotes the time derivative.

Next, we derive the relation for the time derivative of the inflaton field:
\begin{equation}\label{dotphi}
    \dot{\phi} = \sigma \mu^2 \sqrt{\overline{K} - 3F - r} \left( 1 + \frac{\delta X}{X_f} + \cdots \right)^{1/2}.
\end{equation}

To confirm the viability of the slow-roll approximation, i.e., Eq. (\ref{SREOM}), we require $ \dot{K}/HK \ll 1 $, $ \dot{F}/HF \ll 1 $, which naturally corresponds to $ \epsilon \ll 1 $. Additionally, we need $ \dot{r}/Hr \ll 1 $. The sufficient conditions for these requirements can be expressed in terms of the slow-roll parameters:
\begin{equation}\label{SRcondition1}
    \epsilon \ll 1, \quad \eta = \frac{\dot{K}}{HK} = \frac{K_{\phi} u}{H K} \ll 1.
\end{equation}

The crucial slow-roll parameter $ \epsilon $ governs the inflationary dynamics, indicating that when $ \epsilon \ll 1 $, the system remains in the slow-roll regime, while $ \epsilon \rightarrow 1 $ marks the end of inflation. For this model, we have:
\begin{equation}\label{epsilon}
    \epsilon = -\frac{\dot{H}}{H^2} = \frac{6}{3 + \frac{9F + \overline{K}}{\frac{2X}{\mu^4} + 3F - \overline{K} + r}}.
\end{equation}
When the slow-roll conditions hold, $ X_{sr} \cong X_f $, and we have $ \epsilon \ll 1 $. If $ \delta X \sim X_f $, the slow-roll conditions are violated, and inflation terminates.

The characteristic of slow-roll for warm inflation is that radiation production remains quasi-stable: $ \dot{\rho}_r / 4H \rho_r \ll 1 $, or equivalently, $ \dot{s}/Hs \ll 1 $. In the slow-roll regime, we have:
\begin{equation}\label{dotrhor}
    \frac{\dot{\rho}_r}{4H \rho_r} = \frac{\dot{r}}{4H r} + \frac{\dot{u}}{2H u}.
\end{equation}

Thus, the condition for the validity of this relation is $ \dot{u}/Hu \ll 1 $ and $ \dot{r}/Hr \ll 1 $. From the time derivative of the Friedmann equation, we get:
\begin{equation}\label{dotFriedmann}
    \epsilon = -\frac{\dot{H}}{H^2} = \frac{2 \left( \frac{3u^2}{\mu^4} + K + \frac{3}{2}r \right) \dot{u}}{\left( \frac{3u^2}{\mu^4} + 2K + 3r \right) H u} \ll 1,
\end{equation}
which shows that $ \dot{u}/Hu \ll 1 $ is naturally satisfied in the slow-roll regime.

By integrating the Stefan-Boltzmann law $ \rho_r = \sigma T^4 $, we obtain:
\begin{equation}\label{dotr}
    \left( 1 - \frac{c}{4} \right) \frac{\dot{r}}{H r} = \beta + c \frac{\dot{u}}{H u} + \epsilon.
\end{equation}

Given that $\epsilon\ll1$ and $\dot u/Hu\ll1$ are automatically satisfied, the sufficient conditions for ensuring $\dot r/Hr\ll1$ are $\beta\ll1$. When this analysis is combined with the prior discussions and conclusions that necessitate $\dot K/HK\ll1$ and $\dot r/Hr\ll1$, the sufficient conditions required for fulfilling the slow roll criteria are:
\begin{equation}\label{SRconditions2}
\epsilon\ll1,\quad\quad\quad\eta\ll1,\quad\quad\quad\beta\ll1,
\end{equation}
where the first condition is naturally satisfied when the inflation system remains at the slow roll attractor point. Thus, the truly effective slow roll conditions to ensure $\delta X/X_f\ll1$ and $\dot\rho_r/H\rho_r\ll1$ in this new model are primarily $\eta\ll1$ and $\beta\ll1$. These conditions are not particularly challenging to achieve; with the exception of scenarios where both $K$ and $\Gamma$ remain constant, the model is not dynamical, and its implementation can be categorized into three types as shown in the following Table \ref{tab:cases}.
\begin{table}[htbp]
\centering
\caption{Cases for \( K(\phi) \) and \(\Gamma(\phi)\) satisfying slow roll conditions}
\label{tab:cases}
\begin{tabular}{cc}
\hline
\textbf{Case} & \textbf{Formula} \\
\hline
A & both $ K(\phi)$ and $\Gamma(\phi)$ have weak dependence on $\phi$ \\
B & $K = \text{const.}$, while $\Gamma(\phi)$ has weak dependence on $\phi$ \\
C & $\Gamma = \text{const.}$, while $K(\phi)$ has weak dependence on $\phi$ \\
\hline
\end{tabular}
\end{table}

As detailed in Table \ref{tab:cases}, our slow-roll conditions are readily met in several generic cases, and the slow roll condition for $K(\phi)$ and $\Gamma(\phi)$ has a minor dependence on the inflaton field, which is easily satisfied. Some illustrative examples of functions $K(\phi)$ or $\Gamma(\phi)$ that can meet this condition include: (a) any power law or exponential growth as $\phi\rightarrow\infty$, (b) a function dominated by a constant term, such as $K(\phi)=K_0+\lambda (\phi/m)^{\alpha}$, or (c) any leveling-off function of $K(\phi)$ or $\Gamma(\phi)$, i.e., $K(\phi)$ (or $\Gamma(\phi)$) $\rightarrow$ limit, with $K_{\phi}$ (or $\Gamma_{\phi}$) $\rightarrow 0$.

Analyzing the slow roll conditions alongside the examples provided above reveals that, unlike GR-based potential-driven inflationary models where slow-roll requires an extremely flat potential (i.e., the potential slow roll parameters $\epsilon_V=\frac{M_p^2}{2}\frac{V_{\phi}^2}{V^2}, \eta_V=M_p^2\frac{V_{\phi\phi}}{V}\ll 1$), our kinetic-driven attractor demands only a weak field-dependence of the kinetic and dissipative functions. This requirement is inherently simpler to satisfy and decouples inflation from flat potentials, making it possible to achieve a slow roll inflation scenario capable of generating sufficient e-folds without relying on any potential. In addition, the strong damping induced by the nonminimal derivative coupling and thermal dissipation lead to an inherently lower kinetic energy of the inflaton field corresponding to our slow-roll attractor solution, and this configuration guarantees that the inflation can be accurately described within the framework of effective field theory. With the introduction of thermal dissipation, inflation can terminate naturally when the radiation energy density surpasses that of the inflaton field. From these viewpoints, this model demonstrates relative effectiveness and success.

\subsection{Attractor Behavior}

We now investigate the attractor behavior of the inflationary solution. To analyze this, we introduce small perturbations around the attractor solution by expanding the scalar field velocity: $u= u_f + \delta u$, where $\delta u$ denotes a small deviation from the exact attractor solution in this section. In this context, we reexpress the perturbation in terms of $\delta X$ rather than $\delta u$, resulting in $\delta X = u_f \delta u$.

Linearizing the equation of motion Eq. (\ref{dotu}), keeping only terms up to $\mathcal{O}(\delta X)$ and considering $d/dN=\frac1H d/dt$, we obtain:
\begin{equation}
    \frac{d}{dN}(\delta X) \simeq -\lambda \delta X,
\end{equation}
where $\lambda$ is an effective damping coefficient, related to the background quantities of the system:
\begin{equation}
\lambda =\frac{3(2X/\mu^4+K+3F+r)+\frac{12 X}{\mu^4}+\frac{K_\phi u_f}{H}}{6X/\mu^4+K}.
\end{equation}

Using the attractor condition $X_f=\mu^4(\overline{K}-3F-r)/2$, and introducing the slow-roll parameter, we obtain a compact form:
\begin{equation}
\lambda =\frac{12 X_f/\mu^4 + \eta K}{6X_f/\mu^4+K}\simeq \frac{12 X_f/\mu^4}{6X_f/\mu^4+K}.
\end{equation}
The attractor solution is stable provided $\lambda>0$, which requires
\begin{equation}
6X_f/\mu^4 + K > 0.
\end{equation}
Using the expression for $X_f$, this condition leads to
\begin{equation}
\overline{K} > \frac{9F + r}{2}.
\end{equation}
Therefore, the perturbation $\delta X$ decays exponentially as
\begin{equation}
\delta X \propto e^{-\lambda N},
\end{equation}
demonstrating that the kinetic attractor is dynamically stable.

This analytical result establishes the local stability of the attractor.
To further demonstrate its global stability and robustness against initial conditions,
we now turn to a numerical phase-space analysis.

\begin{figure}[t]
\centering
\includegraphics[width=0.4\textwidth]{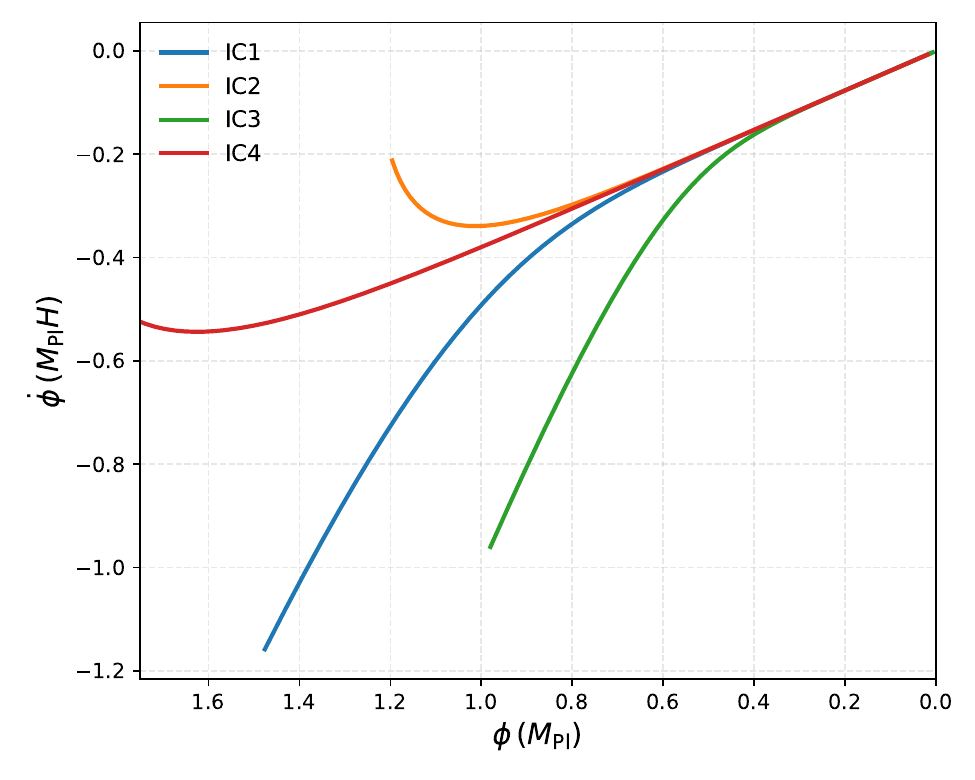}
\caption{
Phase-space trajectories of $(\phi, \dot{\phi})$ obtained from numerical evolution of the background equation. Here $\phi$ is in units of the reduced Planck mass $M_{pl}$, and $\dot{\phi}$ in units of $M_{pl}H$. Different curves correspond to distinct initial conditions (IC1--IC4). All trajectories converge to a common attractor, demonstrating the stability and robustness of the slow-roll solution against initial conditions.
}
\label{fig:phase}
\end{figure}

The phase-space trajectories are obtained by numerically evolving a reduced dynamical system that captures the leading overdamped behavior near the attractor. In the strong dissipation and strong coupling regime, the full equation of motion can be approximated by a linearized system around the attractor solution.
After expanding around the attractor and rescaling variables to dimensionless form, the dynamics can be expressed as
$\frac{d\dot{\phi}}{dN} = -\mathcal{C}_1 \dot{\phi} - \mathcal{C}_2 \phi$, where $\mathcal{C}_1$ and $\mathcal{C}_2$ are positive coefficients determined by the background quantities.
For illustrative purposes, and without loss of generality for demonstrating the attractor behavior, we adopt the representative choice $\mathcal{C}_1=3$ and $\mathcal{C}_2=1$. These coefficients are obtained by evaluating the linearized background equation near the attractor and rescaling variables to dimensionless form; their precise values do not affect the qualitative existence of the attractor. This reduced system captures the essential overdamped dynamics and reproduces the universal attractor behavior of the full system. The four trajectories correspond to the following initial conditions:
\begin{itemize}
    \item \textbf{IC1:} $(\phi_0,\dot{\phi}_0) = (1.5,\,-1.2)$
    \item \textbf{IC2:} $(\phi_0,\dot{\phi}_0) = (1.2,\,-0.2)$
    \item \textbf{IC3:} $(\phi_0,\dot{\phi}_0) = (1.0,\,-1.0)$
    \item \textbf{IC4:} $(\phi_0,\dot{\phi}_0) = (1.8,\,-0.5)$
\end{itemize}

These initial conditions are chosen to span a representative region of phase space, including both large and small initial velocities. Despite their differences, all trajectories rapidly converge to the same attractor curve, confirming that the inflationary solution is dynamically stable and insensitive to initial conditions. This demonstrates that the purely kinetic inflationary solution is not only dynamically stable but also robust against a wide range of initial conditions, establishing it as a genuine attractor of the system.

\subsection{Consistency Analysis of the Pure Kinetic Model}

We now analyze the stability of the model at the level of linear perturbations.
The quadratic action for scalar perturbations can be written in the standard form:
\begin{equation}
    S^{(2)} = \int dt \, d^3x \, a^3
    \left[ \mathcal{G} \dot{\zeta}^2 - \mathcal{F} \frac{(\nabla \zeta)^2}{a^2} \right],
\end{equation}
where $\mathcal{G}$ and $\mathcal{F}$ denote the effective kinetic and gradient coefficients, respectively.
Their explicit forms can be systematically derived from the linear perturbation equation presented in Sec. \ref{sec:level4} (such as Eqs. (\ref{pEOM1}) and (\ref{pEOM2})).
Here we focus on their leading-order expressions relevant for stability analysis. Thus, in the NMDC warm $k$-inflation framework, these coefficients take the forms:
\begin{equation}
\mathcal{G} \simeq \mathcal{L}_X c_s^{-2} + 3F,
\qquad
\mathcal{F} \simeq \mathcal{L}_X + 3F.
\end{equation}

\vspace{0.3em}

The effective sound speed, characterizing the propagation of curvature perturbations in the multi-component warm inflationary system, is defined as
\begin{equation}
    c_{s,\mathrm{eff}}^2 = \frac{\mathcal{F}}{\mathcal{G}}.
\end{equation}

For the purely kinetic Lagrangian $\mathcal{L} = K(\phi) X + \frac{X^2}{\mu^4}$,
the intrinsic inflaton sound speed is therefore $c_s^2 = \frac{K + \frac{2X}{\mu^4}}{K + \frac{6X}{\mu^4}}$.
Substituting into $c_{s,\mathrm{eff}}^2$, we obtain $c_{s,\mathrm{eff}}^2= \frac{K + \frac{2X}{\mu^4} + 3F}
{K + \frac{6X}{\mu^4} + 3F}$.

\vspace{0.3em}

Evaluating this expression at the attractor solution $X_f$, we obtain a compact result:
\begin{equation}
c_{s,\mathrm{eff}}^2
= \frac{r}{6F + 3r-2\overline{K}}.
\end{equation}

\vspace{0.3em}

The stability conditions require:
\begin{equation}
\mathcal{G} > 0, \qquad c_{s,\mathrm{eff}}^2 > 0.
\end{equation}

Since $r>0$, the numerator is positive, and thus the denominator must also be positive.
Combining this requirement with the positivity of the attractor solution $X_f > 0$, we obtain the consistency window:
\begin{equation}
3F + r < \overline{K} < 3F + \frac{3}{2} r.
\end{equation}

\vspace{0.3em}

This condition provides a clear and nontrivial constraint on the parameter space.
Notably, in the strong dissipation regime ($r \gg 1$), the allowed range for $\overline{K}$ becomes sufficiently broad, making the stability conditions naturally satisfied without fine-tuning.

\vspace{0.3em}

\textbf{Remark:} The effective sound speed $c_{s,\mathrm{eff}}^2$ differs from the intrinsic field sound speed due to the presence of both thermal dissipation and NMDC effects.
It characterizes the propagation of curvature perturbations in the coupled inflaton-radiation system and is therefore the physically relevant quantity for assessing gradient stability in warm inflation.

\vspace{0.5em}

\noindent\textbf{Combined constraints on the parameter space.}
By combining the attractor stability condition and the perturbation stability requirements, we obtain a set of nontrivial constraints on the model parameters.

From the attractor analysis, the stability condition requires $\overline{K} > \frac{9F + r}{2}$. On the other hand, the absence of ghost and gradient instabilities imposes $3F + r < \overline{K} < 3F + \frac{3}{2}r$. The effective lower bound is determined by the stronger of the two kinds of frictions, leading to the following consistency windows:

\begin{itemize}
\item For $r \leq 3F$:
\begin{equation}
\frac{9F + r}{2} < \overline{K} < 3F + \frac{3}{2}r,
\end{equation}

\item For $r \geq 3F$:
\begin{equation}
3F + r < \overline{K} < 3F + \frac{3}{2}r.
\end{equation}
\end{itemize}

These results demonstrate that a consistent and stable parameter region exists without fine-tuning.
In particular, in the strong dissipation regime ($r \gg 1$), the allowed window becomes sufficiently broad, ensuring that the model naturally satisfies both dynamical and perturbative stability conditions.

This consistency window provides a robust theoretical foundation for the viability of the purely kinetic NMDC warm inflation scenario. Taken together, these conditions define a nontrivial but well-controlled parameter region
in which the model is both dynamically stable and free of gradient instabilities.

\section{\label{sec:level4} Cosmological Perturbations}

After establishing the background dynamics and the slow-roll attractor, we now calculate the primordial perturbations within the NMDC warm $k$-inflationary framework. This framework differs from cold inflation primarily in the origin of density fluctuations, where the dominant source transitions from quantum fluctuations in cold inflation to thermal fluctuations in warm inflation. While warm inflationary models can theoretically generate both entropy and curvature perturbations, this analysis focuses solely on curvature perturbations. This choice is justified because the subdominant radiation energy mainly contributes to entropy perturbations, which decay on large scales \cite{Ian2008,Cai2011,Lisa2004}, whereas curvature perturbations remain significant.

The inflaton field is extended as $\Phi(\mathbf{x}, t) = \phi(t) + \delta\phi(\mathbf{x}, t)$, where $\phi(t)$ is the homogeneous background field, and $\delta\phi(\mathbf{x}, t)$ represents the linear perturbation caused by thermal stochastic noise $\xi$ in the thermal system. In the high-temperature regime, $T \to \infty$, the thermal noise adopts a Markovian character:
$\langle \xi(\mathbf{k}, t) \xi(-\mathbf{k'}, t') \rangle = 2 \Gamma T a^{-3} (2\pi)^3 \delta^3(\mathbf{k} - \mathbf{k'}) \delta(t - t')$,
as shown in previous studies \cite{Lisa2004, Gleiser1994}. Due to the complexities of non-Markovian noise, along with limited understanding of the early universe dynamics, warm inflation has typically employed the ``no-memory" approximation, treating the thermal noise as Markovian. This approach is validated in \cite{Brahma2022}, where it is shown that non-Markovian corrections to the power spectrum and other observables are negligible. In fact, entanglement-induced rectifications to the power spectrum, spectral index, and its running are of a very small order \cite{Figueroa}. Thus, incorporating this Markovian thermal noise term into the full inflaton field equation (\ref{EOM0}), we obtain the second-order Langevin equation for the field as:
\begin{eqnarray}\label{pEOM1}
    \left( \mathcal{L}_X c_s^{-2} + 3F \right) \ddot{\Phi}(\mathbf{x}, t) + \left[ 3H \left( \mathcal{L}_X + 3F \right) + \Gamma \right] \dot{\Phi}(\mathbf{x}, t) \nonumber \\
    + \left( \mathcal{L}_X \frac{\nabla^2}{a^2} - 3Fw \frac{\nabla^2}{a^2} \right) \delta\phi(\mathbf{x}, t) + K_\phi X = \xi(\mathbf{x}, t).
\end{eqnarray}

Applying Fourier transforms and neglecting the homogeneous background term, we derive the evolution equation for the fluctuation:
\begin{eqnarray}\label{pEOM2}
    (\mathcal{L}_X c_s^2 + 3F) \delta \ddot{\phi}_{\mathbf{k}} + \left[ 3H (\mathcal{L}_X + 3F + r) \right] \delta \dot{\phi}_{\mathbf{k}} \nonumber \\
    + \left( \mathcal{L}_X \frac{k^2}{a^2} - 3Fw \frac{k^2}{a^2} \right) \delta\phi_{\mathbf{k}} + K_\phi \delta X + X K_{\phi\phi} \delta \phi = \xi_{\mathbf{k}}.
\end{eqnarray}

Solving this second-order Langevin equation analytically is challenging.
However, given that the power spectrum is evaluated at the moment of horizon crossing, which occurs during the overdamped, slow-roll regime \cite{LiddleLyth}, in deriving Eq. (\ref{pEOM3}) we have used the overdamped approximation valid in the slow-roll regime and neglected higher-order corrections suppressed by slow-roll parameters.
This approximation transforms Eq. (\ref{pEOM2}) into a first-order equation:

\begin{equation}\label{pEOM3}
    [3H (\mathcal{L}_X + 3F + r)] \delta \dot{\phi_{\mathbf{k}}} + (\mathcal{L}_X + 3F) \frac{k^2}{a^2} \delta \phi_{\mathbf{k}} = \xi_{\mathbf{k}},
\end{equation}
which is much easier to solve in the slow-roll regime.

In warm inflation, scalar perturbations receive contributions from both quantum and thermal fluctuations. In warm inflation, where $T > H$ and $r>0$, the quantum contribution (similar to that in cold inflation) is subdominant compared to thermal fluctuations, which dominate and lead to enhanced field perturbation (see Appendix \ref{app:thermal} for a detailed analysis). Following standard treatments of warm inflation, the inflaton fluctuations are given by:
\begin{equation}
    \delta \phi^2 \simeq \frac{k_F T}{2\pi^2},
\end{equation}
where $ k_F $ is the freeze-out momentum scale determined by the dissipation strength. This expression can be derived using a stochastic approach or equivalently from the Green's function method applied to the Langevin equation for inflaton fluctuations \cite{Berera1995, Berera2000, Chris2009, BereraIanRamos}.

In warm inflationary regime, the thermal contribution dominates over quantum fluctuations, resulting in an enhanced scalar power spectrum. Within the deep slow-roll regime, the solution for the fluctuation is approximately given by:
\begin{eqnarray}\label{phik}
    \delta \phi_{\mathbf{k}}(t) \simeq \frac{1}{3H (\mathcal{L}_X + 3F + r)} \exp\left[-\frac{t - t_0}{\tau(\phi_0)}\right] \int_{t_0}^{t} \exp\left[\frac{t' - t_0}{\tau(\phi_0)}\right] \nonumber \\
    \times \xi\left(\mathbf{k}, t'\right) dt' + \delta\phi(\mathbf{k}, t_0) \exp\left[-\frac{t - t_0}{\tau(\phi_0)}\right],~~~~~~~~~~
\end{eqnarray}
where $ \tau(\phi) = \frac{3H (\mathcal{L}_X + 3F + r)}{(\mathcal{L}_X + 3F) \frac{k^2}{a^2}} $ represents the thermalization timescale. The larger the value of $ k_F $, the faster the mode relaxes. As the universe expands, when the physical wavenumber of the field mode $ \delta \phi(\mathbf{x}, t) $ falls below $ k_F $, it becomes unaffected by thermal noise in a Hubble time \cite{Berera2000, Zhang2014}. The freeze-out scale $k_F$ is determined by the condition
that the relaxation time equals the Hubble time, $\tau \sim H^{-1}$, leading to the relation:
\begin{equation}
    k_F = H \sqrt{\frac{3(\mathcal{L}_X + 3F + r)}{\mathcal{L}_X + 3F}} = H \sqrt{3(1 + Q)},
\end{equation}
where $Q = r/(\mathcal{L}_X + 3F)$ is the effective dissipative ratio, specific to the NMDC warm $k$ framework, and measures whether thermal effects dominate over other damping mechanisms like NMDC and noncanonical effects. In the strong dissipative regime $ Q \gg 1 $, $ k_F $ exceeds the Hubble rate, indicating that freeze-out occurs before horizon crossing. In contrast, in the weak dissipative regime $ Q \ll 1 $, $ k_F $ is slightly greater than $ H $.

Using the freeze-out condition $k = k_F$, evaluating the fluctuation amplitude at horizon crossing, and applying the spatially flat gauge with $ \mathcal{R} = \frac{H}{\dot{\phi}} \delta \phi_{\mathbf{k}} $, the scalar power spectrum for the NMDC warm $k$-inflationary model can be finally obtained:
\begin{eqnarray}\label{PR}
    P_R = \left( \frac{H}{\dot{\phi}} \right)^2 \delta \phi^2 = \frac{H^3 T}{4 \pi^2 X} \sqrt{3(1 + Q)} \nonumber \\
    = \frac{H^3 T \sqrt{3(1 + Q)}}{2 \pi^2 \mu^4 \left[ \overline{K}(\phi) - 3F - r \right]}.
\end{eqnarray}
This result reduces to the standard warm inflation expression
$P_R \propto HT\sqrt{1+r}$ in the canonical limit, up to model-dependent modifications encoded in noncanonical kinetic structure and NMDC (i.e. $\mathcal{L}_X$ and $F$). It is evident from Eq. (\ref{PR}) that the scalar power spectrum in the NMDC warm $k$-inflation is enhanced by the factor $\frac TH\sqrt{1+Q}$, which explicitly demonstrates the dominance of thermal fluctuations over quantum fluctuations. For completeness, the quantum contribution is of order $\delta\phi^2 \sim (H/2\pi)^2$, which is completely subdominant compared to the thermal contribution, given that in warm inflation the relations $T > H$ and $(1+Q)>1$ are always satisfied.

The scalar power spectrum is typically normalized to $ P_R \approx 10^{-9} $ on large scales, as indicated by Cosmic Microwave Background (CMB) observations. The factor $\sqrt{1 + Q}$ in the numerator suggests that the energy scale at horizon crossing can be considerably reduced due to thermal dissipation. This reinforces the assumption that inflation can be effectively described within the field theory framework.

The expression for the spectral index $n_s$ is obtained by substituting the slow-roll solutions into Eq. (\ref{PR}) and evaluating the resulting power spectrum and its derivatives at horizon crossing, retaining only leading-order terms:
\begin{eqnarray}
    n_s - 1 = \frac{d \ln P_R}{d \ln k} \simeq \frac{\dot{P_R}}{H P_R} \nonumber \\
    = \alpha_1 \epsilon + \alpha_2 \eta + \alpha_3 \beta.
\end{eqnarray}
The parameters $ \alpha_1, \alpha_2, \alpha_3 $ in above equations are defined as:
\begin{eqnarray}
    \alpha_1 &=& - \frac{K}{2(9F - K)} \left[ 3 + \frac{Q \mathcal{L}_X (c_s^{-2} - 1)}{(1 + Q)(\mathcal{L}_X + 3F)} \right] - 3 \nonumber \\
    &+& \frac{4}{4 - c} \left[ 1 + \frac{r}{2(1 + Q)(\mathcal{L}_X + 3F)} \right] \left[ 1 + \frac{c K}{2(9F - K)} \right] \nonumber \\
    &+& \frac{3F Q}{(1 + Q)(\mathcal{L}_X + 3F)},
\end{eqnarray}
\begin{eqnarray}
    \alpha_2 &=& \frac{K}{4(9F - K)} \left\{ 3 + \frac{Q \mathcal{L}_X (c_s^{-2} - 1)}{(1 + Q)(\mathcal{L}_X + 3F)} \right. \nonumber \\
    &-& \left. \frac{4c}{4 - c} \left[ 1 + \frac{r}{2(1 + Q)(\mathcal{L}_X + 3F)} \right] \right\},
\end{eqnarray}
\begin{eqnarray}
    \alpha_3 = \frac{4}{4 - c} \left[ 1 + \frac{r}{2(1 + Q)(\mathcal{L}_X + 3F)} \right].
\end{eqnarray}

These parameters are of order unity, so $ (n_s - 1) $ is a first-order small value in the slow-roll regime, specifically of order $ \epsilon $. Therefore, the model predicts a nearly scale-invariant power spectrum, which aligns well with observations. The running of the spectral index $ \alpha_s = \frac{dn_s}{d \ln k} $ is estimated to be on the order of $ \mathcal{O}(\epsilon^2) $, which is much smaller than $(n_s - 1) $, further corroborating the observational agreement.

Tensor perturbations are not affected by the thermal background and arise only from quantum fluctuations, similar to standard inflation \cite{TalyorBerera}. Only a minor rectification is made by the nonminimal derivative coupling $G^{\mu\nu} \partial_{\mu}\phi\partial_{\nu}\phi$ to the power spectrum of gravitational waves \cite{DalianisJCAP2020,Karydas2102.08450}:
\begin{equation}\label{tensorperturbation}
P_T\simeq\frac{2}{M_p^2}\left(\frac{H}{2\pi}\right)^2.
\end{equation}
Through first-order approximation, the spectral index of tensor perturbation is given by $n_T=-2\epsilon$. In this model, the tensor-to-scalar ratio $R$ is represented as:
\begin{equation}\label{R}
R=\frac{P_{T}}{P_{R}}=\frac{H}{T}\frac{12}{(9F+\overline{K})\sqrt{3(1+Q)}}<<1.
\end{equation}
Since the scalar power spectrum is fixed by observations, the impacts induced by nonminimal derivative coupling and thermal dissipation considerably reduce tensor perturbations. Therefore, a strong NMDC impact (i.e., $F\gg1$) along with significant thermal dissipation (i.e., $Q\gg1$) can both lead to a notable decrease in the magnitude of gravitational wave. This combined suppression leads a small $R$ to be a natural prediction of the framework, which is consistent with observations without fine-tuning. The influence of the strong NMDC effect is more pronounced compared to that of the thermal effect to some extent. Furthermore, by applying the criteria for the onset of warm inflation $T>H$, an upper limit for the tensor-to-scalar ratio is obtained:
\begin{equation}
R<\frac{12}{(9F+\overline{K})\sqrt{3(1+Q)}}.
\end{equation}
The contribution to the best-fit tensor-to-scalar ratio corresponds to a value near the observational limit $R=(3-6)\times 10^{-3}$ \cite{BICEP2,PLANCK1}, with the upper bound of $R$ being sufficiently low to meet existing constraints. The results indicate that a very small value of $R$ in this NMDC warm $k$-inflationary model aligns well with observations.

The negligible non-Gaussianity suggested by {\it Planck} observations \cite{PLANCK2} also supports a relatively high sound speed and moderate thermal dissipation \cite{Cai2011,Franche2010}. The magnitude and shape of non-Gaussianity in our new model are expected to be qualitatively influenced by key physical quantities, i.e., $f_{NL}$ relies on $(c_s^{-2},F,Q)$. A more thorough investigation into non-Gaussian characteristics and the observational constraints of non-Gaussianity concerning the parameters of the NMDC warm $k$-inflation model, particularly the significant impact of purely kinetic characteristics, merits further exploration and will be considered by our forthcoming research.

The expansion amount is $\Delta N\simeq 4.6$ when the scales corresponding to $1\leq l \leq100$ exits the horizon, with the field variation given by:
\begin{equation}
\frac{\Delta\phi}{M_{p}}=\frac{\dot{\phi}\Delta N}{M_p H}\simeq\frac{9.2\sqrt{3}}{\sqrt{9F+\overline{K}}}\sim(9F+\overline{K})^{-\frac12}.
\end{equation}
In the NMDC warm $k$-inflationary model, the standard Lyth bound in the GR frame \cite{LythPRL1997} is modified, revealing that the field excursion is constrained to be less than the Planck scale ($\Delta\phi\ll M_p$), particularly in scenarios with high gravitational friction. This contrasts with standard inflation ($\frac{\Delta\phi}{M_p}=0.5(\frac{R}{0.1})^{1/2}$, and a detectable $R$ necessitates Planckian or even super-Planckian inflation excursions \cite{LiddleLyth}). Consequently, the problem of a too large inflaton amplitude is effectively tackled in this novel NMDC warm $k$-inflationary model, even without any potential term. As shown by Eq. (\ref{R}), the consistency relation is no longer fixed, significantly differing from the scenario of minimal cold inflation ($R=-6.2n_T$ \cite{LiddleLyth}).

As previously mentioned, the radiation energy density and the universal temperature  adhere to the Stefan-Boltzmann relationship $\rho_r=\pi^2g_{\ast}T^4/30$. The integration of the radiation production slow-roll approximation and the power spectrum yields:
\begin{equation}\label{TH}
\frac{T}{H}=\left(\frac{45r}{4\pi^4g_*P_R}\right)^{\frac{1}{3}}[3(1+Q)]^{\frac{1}{6}}.
\end{equation}
$T/H$ is marginally lower than that observed in warm inflationary cases in the GR limit \cite{Ian2008,Zhang2014}. Nonetheless, a higher thermal dissipation strength $Q$ leads to a greater $T/H$ ratio. In situations where strong thermal dissipation prevails over gravitational damping $Q\gg1$, the temperature of the universal bath increases. In contrast, weak dissipation results in a temperature drop because the thermal effect contends with the NMDC effect. Since the factor $(1+Q)$ is always greater than one, the condition for the onset of warm inflation, $T>H$, can be adequately met when $r\geq g_{\ast}P_R$, as demonstrated by Eq. (\ref{TH}). Since $g_{\ast} \sim \mathcal{O}(10^2)$ and $P_R \sim \mathcal{O}(10^{-9})$ during inflation, the threshold for initiating warm inflation is notably low. In fact, even minimal thermal dissipation, requiring only $r \geq 10^{-7}$, is sufficient. It should be noted that in our NMDC warm inflation framework, the newly defined parameter $Q = r / (\mathcal{L}_X + 3F)$, which is more appropriate for this framework than the traditional $r$, is actually less than $r$. Thus, warm inflation could more realistically describe the very early accelerating universe. In addition, even in the weak dissipative regime $Q\ll1$, thermal fluctuations still play a significant role in the perturbations, which is quite different from cold $k$-inflationary scenarios, as a direct consequence of Eq. (\ref{PR}).

This confirms that the present NMDC warm $k$-inflation framework provides a self-consistent and observationally viable realization of inflation driven purely by kinetic terms.

\section{\label{sec:level5} Concrete Model and Observational Constraints}

In this section, we present a concrete realization of the NMDC warm $k$-inflationary model and confront its predictions with observational data, particularly from the Planck 2018 survey \cite{PLANCK1}.

To explicitly demonstrate the phenomenological viability of the model, we construct a concrete realization:
\begin{equation}
K(\phi) = -K_0 \left(1 + \alpha \phi \right),
\end{equation}
where $K_0$ and $\alpha$ are positive phenomenological constants. We assume a constant dissipation coefficient for simplicity: $\Gamma = \Gamma_0$, which adequately captures the essential features of the warm inflationary regime.

To quantitatively disentangle the interplay among nonminimal derivative coupling, thermal dissipation, and noncanonical kinetic effects, we focus on the key dimensionless quantities that directly enter the background and perturbation dynamics:
\begin{equation}
F \equiv \frac{H^2}{M^2}, \qquad
r \equiv \frac{\Gamma}{3H}, \qquad
\mathcal{L}_X = K(\phi) + \frac{2X}{\mu^4}.
\end{equation}
Here, $F$ characterizes the strength of the NMDC-induced gravitational friction,
$r$ measures the thermal dissipation rate, and $\mathcal{L}_X$ controls the effective
kinetic structure of the inflaton field. In particular, deviations of $\mathcal{L}_X$ from unity encode the importance of noncanonical kinetic effects.

\vspace{0.3em}

Depending on the relative magnitudes of $(F, r, \mathcal{L}_X)$, the system can be qualitatively classified into four dynamical regimes:
\begin{itemize}
\item \textbf{NMDC-dominated regime:} $F \gg r, \mathcal{L}_X$,
where gravitational friction provides the dominant damping. Thus, the model essentially returns to the NMDC-driven canonical cold inflation.

\item \textbf{Thermal-dominated regime:} $r \gg F, \mathcal{L}_X$,
where dissipation controls both the background evolution and fluctuations. Thus, the model almost returns to the standard warm inflation with a canonical scalar field.

\item \textbf{Kinetic-dominated regime:} $\mathcal{L}_X$ deviates significantly from unity, $r, F \ll 1$,
where higher-order kinetic terms dominate the dynamics, reducing the model to pure kinetic cold inflation.

\item \textbf{Mixed regime (this work):} $\mathcal{L}_X \gtrsim 1$, $r \sim 1$, $F \sim 1$,
where thermal dissipation, NMDC, and noncanonical kinetic effects are all of comparable strength, leading to a cooperative interplay that renders the inflationary dynamics particularly robust and consistent with observations.
\end{itemize}
The first three regimes correspond to limiting corners of the parameter space, where a single effect overwhelmingly dominates. In contrast, a large and phenomenologically relevant portion of the parameter space, including the region favored by observational data, belongs to the mixed regime, in which all three effects contribute comparably. These four regimes are summarized in Table~\ref{tab:regimes}.

Based on the parameter space obtained from the analysis of attractor stability and model self-consistency in Sec.~\ref{sec:level3}, we evaluate the background solution and find that the three contributions are typically of the same order, with a mild hierarchy given by
\begin{equation}
r \gtrsim F \gtrsim \mathcal{L}_X.
\end{equation}

We emphasize that the hierarchy above reflects the typical relative ordering of these dimensionless quantities across the broad parameter space of the NMDC warm $k$-inflation framework, without fixing their absolute magnitudes. The subsequent comparison with Planck data is performed within a specific, simplified realization of the theory---namely the linear kinetic function $K(\phi)=-K_0(1+\alpha\phi)$ with a constant dissipation coefficient---and the resulting numerical ranges for $r$ are therefore model-dependent. Nevertheless, the qualitative hierarchy remains a robust feature of the underlying dynamics.

This quantitative characterization clarifies the interplay among the three physical mechanisms and demonstrates that the observationally viable region of the model does not rely on a single dominant effect, but rather on their combined action.
\begin{table}[t]
\centering
\begin{tabular}{c|ccc}
\hline
Regime & $\mathcal{L}_X$ & $r$ & $F$ \\
\hline
NMDC-dominated & $\sim1$ & $\ll1$ & $\gg 1$ \\
Thermal-dominated & $\sim1$ & $\gg 1$ & $\ll1$ \\
Kinetic-dominated & $> 1$ & $\ll 1$ & $\ll 1$ \\
Mixed (this work) & $\gtrsim 1$ & $\sim 1$ & $\sim 1$ \\
\hline
\end{tabular}
\caption{Characterization of the different dynamical regimes based on the relative strengths of NMDC, thermal dissipation, and noncanonical kinetic effects.}
\label{tab:regimes}
\end{table}
These parameters provide a quantitative understanding of the interplay among the various mechanisms in the model.

To illustrate the background dynamics in a concrete setting, we numerically solve the slow-roll system for the specific model introduced above. Instead of evolving the full stiff system, we focus on the attractor regime, which governs the physically relevant dynamics around horizon crossing.

\begin{figure}[t]
\centering
\includegraphics[width=0.4\textwidth]{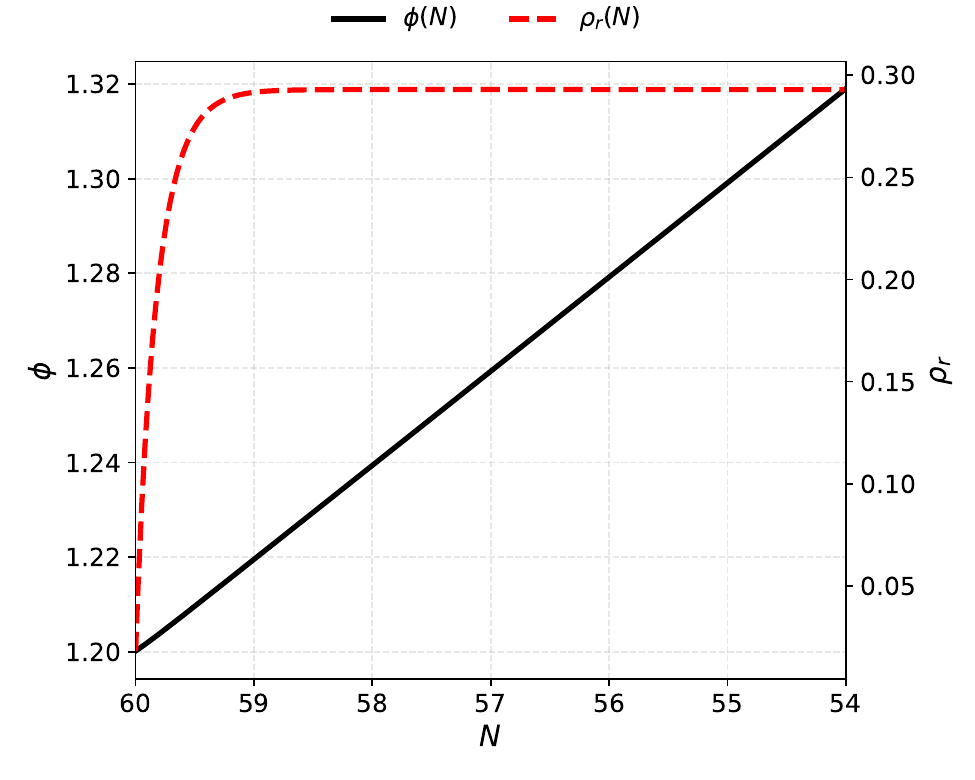}
\caption{
Numerical evolution of the background quantities as functions of the number of
e-folds $N$. The scalar field $\phi$ evolves smoothly along the attractor trajectory,
while the radiation energy density $\rho_r$ remains subdominant but non-negligible.
This behavior is characteristic of the warm inflation regime, where radiation is
continuously produced but does not dominate the total energy density.
}
\label{fig:evolution}
\end{figure}

Figure \ref{fig:evolution} shows the evolution of the inflaton field $\phi$ and the radiation energy density $\rho_r$ as functions of the number of e-folds $N$ in the interval $N \in [54,60]$, corresponding to the epoch when cosmological scales exit the horizon. Restricting to this window allows us to directly probe the slow-roll attractor behavior relevant for observable predictions. We observe that $\rho_r$ quickly grows from negligible values and then approaches a quasi-stationary plateau. This behavior reflects a balance between radiation production and Hubble dilution, leading to a sustained but subdominant radiation component. Such a quasi-steady state is a characteristic feature of warm inflation and confirms that thermal dissipation remains active throughout the inflationary stage. Meanwhile, $\phi(N)$ evolves approximately linearly in $N$, indicating that $\dot{\phi}$, and hence the kinetic term $X$, remain nearly constant along the trajectory. This is consistent with the kinetic attractor solution derived analytically, where the dynamics are governed by an effective balance between Hubble friction, NMDC-induced gravitational damping, and thermal dissipation.

Overall, the numerical results support the validity of the slow-roll attractor solution and provide a concrete realization of the warm inflation regime in the presence of nonminimal derivative coupling. This behavior also justifies the use of the slow-roll approximation adopted in the perturbation analysis.

\subsection{Model Comparison with Observational Constraints}

In this section, we confront the NMDC warm $k$-inflation model with the Planck~2018 CMB temperature and polarization data \cite{PLANCK1}. The analysis is based on the explicit model introduced in this section, namely $K(\phi)=-K_0(1+\alpha\phi)$ with a constant dissipation coefficient $\Gamma=\Gamma_0$. At the level of the slow-roll approximation, the scalar spectral index $n_s$ and the tensor-to-scalar ratio $R$ are controlled by three effective parameters: the NMDC strength $F \equiv H^2/M^2$, the kinetic factor $\mathcal{L}_X \equiv \partial \mathcal{L}/\partial X$, and the dissipative ratio $r \equiv \Gamma/(3H)$. The stability and consistency conditions derived in Sec. \ref{sec:level3} are enforced throughout.

\begin{figure}[t]
\centering
\includegraphics[width=0.4\textwidth]{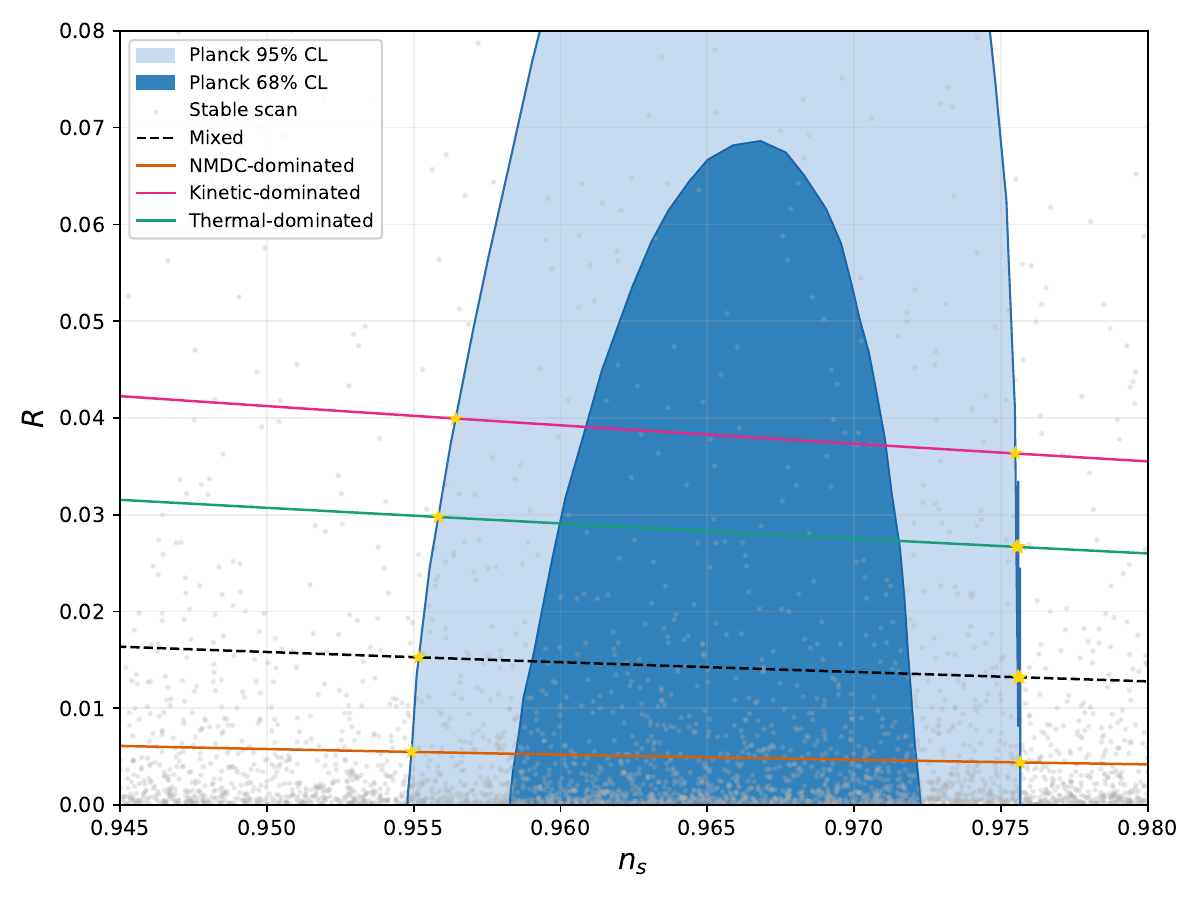}
\caption{
Comparison of the NMDC warm $k$-inflation model with Planck 2018 constraints \cite{PLANCK1}.
Blue shaded regions denote the $68\%$ and $95\%$ confidence level contours.
Light gray dots represent random parameter samples satisfying all stability conditions (Sec.~\ref{sec:level3}) over $F\in[10^{-2},10^3]$, $\mathcal{L}_X\in[0.1,1.5]$, $r\in[10^{-2},10^3]$.
Colored curves show four representative trajectories, each illustrating a regime where the corresponding physical effect is dominant:
\textbf{mixed} (black dashed, $F=10.0$, $\mathcal{L}_X=1.10$);
\textbf{NMDC-dominated} (orange solid, $F=30.0$, $\mathcal{L}_X=1.05$);
\textbf{kinetic-dominated} (purple solid, $F=5.0$, $\mathcal{L}_X=1.30$);
\textbf{thermal-dominated} (green solid, $F=5.0$, $\mathcal{L}_X=1.05$).
For each trajectory, $r$ varies over $[0.01,1000]$.
Golden stars mark the intersections with the $95\%$ confidence boundary, indicating the observationally allowed range of $r$.
}
\label{fig:nsr}
\end{figure}

To assess the observational viability of the model, two complementary approaches are employed, as summarized in Fig. \ref{fig:nsr}. First, a set of representative trajectories is constructed by fixing $F$ and $\mathcal{L}_X$ at selected values, while continuously varying the dissipation parameter $r$ over $[0.01,100]$.
These trajectories, plotted as colored curves, illustrate how the predicted $(n_s,R)$ evolve with the thermal dissipation strength for given gravitational and kinetic backgrounds.
The intersections of the trajectories with the $95\%$ confidence boundary are marked by golden stars, delimiting the observationally allowed range of $r$ for each $(F,\mathcal{L}_X)$ pair.

Second, a dense random sampling is performed over the broader parameter ranges $F \in [10^{-2},10^3]$, $\mathcal{L}_X \in [0.1,1.5]$, and $r \in [10^{-2},10^3]$, with the resulting points displayed as light gray dots.
This sampling reveals the global projection of the full three-dimensional parameter space onto the $(n_s,R)$ plane and confirms that the agreement with observational constraints is not confined to isolated fine-tuned trajectories but holds across a wide and natural region of the model's parameter space.

Figure~\ref{fig:nsr} presents the observational constraints on the NMDC warm $k$-inflation model in the $(n_s,R)$ plane. The light gray points represent a dense scan over the parameter space $(F,\mathcal{L}_X,r)$ restricted by the attractor stability, ghost-free, and gradient stability conditions established in Sec.~\ref{sec:level3}.
A substantial fraction of these points fall within the Planck $68\%$ and $95\%$ confidence regions, demonstrating that the agreement with data is not the result of fine-tuning but rather a robust feature of the model. A notable feature is that the observationally allowed region is predominantly populated by configurations in which the three effects-thermal dissipation, NMDC, and noncanonical kinetic-are of comparable strength, providing further support for the dynamically realized mixed regime.

Four characteristic trajectories are highlighted, each chosen to illustrate a distinct balance among the three physical effects. For the three ``-dominated'' trajectories, the corresponding effect is maximized by setting its characteristic quantity to a value that makes it the strongest among the three, representing limiting corners of the parameter space; in contrast, the mixed regime covers a much broader and more natural region. In the mixed regime ($F=10.0$, $\mathcal{L}_X=1.10$, black dashed curve), the contributions from NMDC, thermal dissipation, and noncanonical kinetics are of comparable importance, resulting in the allowed dissipation parameter $r\in[0.07,0.09]$. The NMDC-dominated trajectory ($F=30.0$, $\mathcal{L}_X=1.05$, orange solid curve) exhibits stronger gravitational friction and yields a similar allowed interval $r\in[0.09,0.12]$, with the tensor-to-scalar ratio suppressed to $R\sim5\times10^{-3}$. The kinetic-dominated case ($F=5.0$, $\mathcal{L}_X=1.30$, purple solid curve) favors a smaller dissipation window $r\in[0.03,0.04]$ and produces the largest $R$ among the four regimes, reflecting the reduced damping when noncanonical kinetic terms provide the dominant contribution.

Finally, the green solid curve ($F=5.0$, $\mathcal{L}_X=1.05$) is intended to represent the thermal-dominated regime, where by construction both the NMDC and noncanonical kinetic effects are set to be relatively weak. In this limit, one would expect thermal dissipation to control the dynamics. However, the observationally allowed range for this trajectory is $r\in[0.07,0.08]$, which is not large enough for thermal effects to genuinely dominate; instead, it falls within the mixed regime. We therefore refer to this as a \textit{pseudo-thermal-dominated} case, where the thermal effect remains comparable to the other two. It yields an intermediate level of tensor suppression. It is worth noting that the allowed $r$ interval in this thermal-dominated case is not exceptionally large. This is because, for the chosen parameters, a further increase in $r$ would shift the spectral index $n_s$ away from the central Planck value, moving the trajectory outside the $95\%$ confidence region. In other words, while thermal dissipation efficiently suppresses the tensor-to-scalar ratio, the precise location of the trajectory in the $(n_s,R)$ plane is sensitive to the combined effects of $F$, $\mathcal{L}_X$, and $r$. Consequently, both the CMB data and the theoretical stability and consistency analysis favor a mixed regime, in which thermal dissipation plays a significant but not exclusive role, while NMDC and noncanonical kinetic effects provide the additional friction required for full compatibility with observations.

All four trajectories consistently predict $n_s\approx0.965$ within their respective favored $r$ windows. This analysis confirms that the NMDC warm $k$-inflation scenario accommodates current CMB data across a wide range of physically distinct regimes, without requiring any scalar potential. This further indicates that the observational viability is a structural feature of the model rather than an artifact of parameter tuning.

A more detailed numerical exploration of specific benchmark models, together with a full Bayesian parameter estimation using cosmological likelihoods, will be presented in a forthcoming work.

\section{\label{sec:level6}Conclusions}

\subsection{Systematic Comparison with Potential-Driven Inflation}

This work demonstrates that a purely kinetic inflationary model, without invoking any potential term, can simultaneously achieve stable attractor dynamics, consistent perturbation spectra,
and excellent agreement with observations. To clearly distinguish the physical mechanism of this study from the potential-driven scenario discussed in Ref. \cite{Zhang2025}, we summarize the key differences in background dynamics and observational predictions in Table \ref{tab:modelcomparison}. While both models operate within the framework of NMDC warm inflation, the driving mechanism and stability criteria differ significantly.

\begin{widetext}

\begin{table}[htbp]
\centering
\caption{Key distinctions between potential-driven and potential-free (purely kinetic) warm inflation with nonminimal derivative coupling}
\label{tab:modelcomparison}
\renewcommand{\arraystretch}{1.8}
\small
\begin{tabular}{|>{\centering\arraybackslash}m{0.18\textwidth}
 |>{\centering\arraybackslash}m{0.4\textwidth}|>{\centering\arraybackslash}m{0.4\textwidth}|}
\hline
  \textbf{Aspects} & \textbf{Ref. \cite{Zhang2025}: Potential-driven model } & \textbf{This work: Purely kinetic (potential-free) model} \\
\hline
\textbf{Lagrangian density} &
$\displaystyle \mathcal{L} = K(X) - V(\phi)$
& $\displaystyle \mathcal{L} = K(\phi)X + \frac{X^2}{\mu^4}$ \quad (no $V\left(\phi\right)$ ) \\
\hline
\textbf{Slow-roll parameters} &
$\displaystyle \epsilon = \frac{M_p^2}{2}\left(\frac{V_\phi}{V}\right)^2$, \quad
 $\eta = M_p^2\frac{V_{\phi\phi}}{V}$, \quad
 $\beta = M_p^2\frac{V_{\phi}\Gamma_{\phi}}{V\Gamma}$
& $\displaystyle \epsilon = -\frac{\dot{H}}{H^2}$, \quad
$ \eta = \frac{K_\phi\dot\phi}{H K}$, \quad
$ \beta = \frac{\Gamma_\phi\dot\phi}{H\Gamma}$ \\
\hline
\textbf{Slow-roll requirement} &
Sufficiently flat potential, etc. \par
$\displaystyle\left(\epsilon\ll\mathcal{L}_X+3F+r, \quad \eta\ll\mathcal{L}_{X}c_s^{-2}+3F,\right.$ \par
$\left. \beta\ll\frac{(\mathcal{L}_X c_s^{-2}+3F)(\mathcal{L}_X+3F+r)}{r}\right)$
& Weak field-dependence of $K(\phi)$ and $\Gamma(\phi)$ \par
($\epsilon\ll1$ is naturally satisfied in this scenario,  \par and we only need $\eta \ll 1$ and $\beta \ll 1$) \\
\hline
\textbf{Attractor solution} &
$\displaystyle 3H(\mathcal{L}_X+3F+r)\dot{\phi} +V_\phi=0$
& $\displaystyle X_f = \mu^4\!\left(\frac{\overline{K}}{2} - \frac{3F+r}{2}\right)$ \par (derived from $\rho+p=0$) \\
\hline
\textbf{Scalar power spectrum} &
$P_R \propto V_\phi^{-2}$ \par (largely determined by potential slope)
& $P_R \propto X_f^{-1}$ \par (set by kinetic attractor, independent of $V_\phi$) \\
\hline
\textbf{Tensor-to-scalar ratio} &
$\displaystyle R \sim \frac{H}{T} \frac{\epsilon(\mathcal{L}_X+3F)^{1/2}}{(\mathcal{L}_X+3F+r)^{5/2}}$ \par
(suppressed by both thermal and NMDC effects.)
& $\displaystyle R \sim \frac{H}{T} \frac{1}{(9F+\overline{K})\sqrt{3(1+Q)}}$ \par
(extra suppression from $F$, $\overline{K}$, and $Q$) \\
\hline
\textbf{Field excursion} &
$\displaystyle \frac{\Delta\phi}{M_p} \lesssim (\mathcal{L}_X+3F+r)^{-1/2}$ \par
(sub-Planckian field excursion due to synergistic enhanced friction.)
& $\displaystyle \frac{\Delta\phi}{M_p} \sim (9F+\overline{K})^{-1/2}$ \par
(inherently sub-Planckian field excursion even for moderate $F$.) \\
\hline
\textbf{Theoretical novelty} &
NMDC and thermal dissipation collaboratively relax the potential flatness in warm inflation
& \textbf{First unification} of pure $k$-inflation, warm bath, and NMDC; \par
provides a new inflationary mechanism without a potential. \\
\hline
\end{tabular}
\end{table}

\end{widetext}

As shown in Table \ref{tab:modelcomparison}, a key difference lies in the stability of the slow-roll attractor and the generation of the scalar power spectrum. The slow-roll attractor in the potential-free scenario is maintained by the slowly-varying kinetic and thermal dissipative functions, in stark contrast to the potential-dominated slow-roll of traditional models. Furthermore, in this potential-free framework, density perturbations are driven solely by the kinetic attractor $X_f$, independent of any potential slope. This shift represents a departure from traditional models, where the potential slope, $V_\phi$, is fine-tuned for successful inflation. Our model, however, requires only a weak field dependence for the kinetic function $K(\phi)$ and the dissipation coefficient $\Gamma(\phi)$, making it less stringent and more versatile.

Compared with conventional warm inflation models, the combined effects of NMDC and noncanonical kinetics
significantly enlarge the viable parameter space and naturally suppress the tensor-to-scalar ratio. This dual suppression-due to both the NMDC coupling $F$ and the kinetic function $\overline{K}$-results in a tensor-to-scalar ratio that naturally adheres to the sub-Planckian field excursion constraints, a feature not found in potential-driven models.

\subsection{Discussions and Conclusions}

This study introduces a unified inflationary framework that consistently integrates three essential ingredients: a potential-free, purely kinetic inflaton sector; the thermal bath characteristic of warm inflation; and the enhanced gravitational friction induced by nonminimal derivative coupling. Within this setup, a quasi-exponential expansion is sustained without invoking any scalar potential. The combined effects of the noncanonical kinetic structure, NMDC-enhanced friction, and thermal dissipation lead to a strongly overdamped inflaton dynamics, giving rise to a robust slow-roll attractor.

The slow-roll conditions in this model require only a weak field dependence of the kinetic function $K(\phi)$ and the dissipation coefficient $\Gamma(\phi)$, in stark contrast to the stringent flatness condition on the scalar potential required in standard inflation. Consequently, compared with conventional potential-driven inflation, the slow-roll requirements in our scenario are substantially less restrictive. This feature naturally enlarges the theoretically viable parameter space and alleviates the issue of super-Planckian field excursions, keeping the inflaton evolution inherently sub-Planckian without any fine-tuning.

The stability of this attractor has been rigorously analyzed. Linear perturbations around the attractor solution decay exponentially, with a damping coefficient that remains positive under the condition $\overline{K} > (9F+r)/2$, confirming the dynamical robustness of the kinetic attractor. Moreover, a complete stability analysis of scalar perturbations has been performed. From the quadratic action for curvature perturbations, we derived the effective sound speed $c_{s,\mathrm{eff}}^2 = r/(6F+3r-2\overline{K})$, which is strictly positive within the consistency window $3F+r < \overline{K} < 3F+3r/2$. This window simultaneously guarantees the absence of ghost and gradient instabilities, and together with the attractor condition defines a well-controlled parameter region where the model is both dynamically and perturbatively stable.

The perturbation analysis shows that the model predicts a nearly scale-invariant scalar power spectrum, with the spectral index and its running fully consistent with current cosmological observations. A distinctive prediction is the strong suppression of the tensor-to-scalar ratio $R$, which arises from the interplay between thermal damping and NMDC-induced gravitational friction. This mechanism naturally drives $R$ deep into the observationally allowed region, typically yielding $R \lesssim 10^{-3}$ without parameter fine-tuning.

Numerical evolution of the background confirms the existence of a stable attractor, characterized by a quasi-stationary radiation component and an approximately constant kinetic sector. Direct comparison with Planck 2018 data demonstrates that the model predictions for $(n_s,R)$ fall well within the $68\%$ and $95\%$ confidence regions over a broad parameter range. The observationally favored region corresponds to a mixed regime where NMDC, thermal dissipation, and noncanonical kinetic effects contribute comparably, distinguishing the present scenario from conventional warm inflation where either thermal or potential effects typically dominate. The observational viability is therefore a structural feature of the model rather than an artifact of parameter tuning.

Overall, this work establishes NMDC warm $k$-inflation as a compelling and viable alternative to traditional potential-driven inflationary models. Future work will focus on more detailed numerical exploration of specific benchmark realizations, the investigation of higher-order statistics such as non-Gaussianity, and a full likelihood-based statistical analysis using cosmological data. These developments will further clarify the predictive power and observational signatures of this potential-free inflationary paradigm.

\acknowledgments This work was supported by the Shandong Provincial Natural Science Foundation (Grant No. ZR2021MA037 and No. ZR2022JQ04), the National Natural Science Foundation of China (Grant No. 12575134 and No. 11605100), and the Natural Science Foundation of Henan (Grant No. 232300421351).
\appendix
\section{Thermal Fluctuations, Freeze-out, and Comparison with Standard Warm Inflation}
\label{app:thermal}

In this appendix, we clarify the structure of inflaton fluctuations and explicitly separate thermal and quantum contributions. We further demonstrate that the resulting dynamics are a controlled extension of standard warm inflation.

\vspace{0.5em}

\subsection{Thermal vs Quantum Fluctuations}

Inflaton fluctuations decompose as
\begin{equation}
\delta\phi = \delta\phi_{\rm q} + \delta\phi_{\rm th},
\end{equation}
where quantum fluctuations satisfy
\begin{equation}
\delta\phi_{\rm q}^2 \simeq \left(\frac{H}{2\pi}\right)^2,
\end{equation}
while thermal fluctuations arise from stochastic noise obeying
\begin{equation}
\langle \xi(\mathbf{k}, t)\xi(\mathbf{k}', t') \rangle
= 2\Gamma T a^{-3}(2\pi)^3 \delta^3(\mathbf{k}-\mathbf{k}')\delta(t-t').
\end{equation}

The thermal variance is
\begin{equation}
\delta\phi_{\rm th}^2 \simeq \frac{k_F T}{2\pi^2}.
\end{equation}

The relative magnitude is
\begin{equation}
\frac{\delta\phi_{\rm th}^2}{\delta\phi_{\rm q}^2}
\sim \sqrt{3(1+Q)}\frac{T}{H},
\end{equation}
which is generically $\gg 1$ in the warm regime ($T>H$, $Q\gtrsim 1$), establishing thermal dominance.

\vspace{0.5em}

\subsection{Effective Langevin Structure}

At horizon crossing and in the overdamped regime, the perturbation equation reduces to
\begin{equation}
\gamma_{\rm eff}\,\delta \dot{\phi}_{\mathbf{k}} + \omega_k^2 \delta \phi_{\mathbf{k}} = \xi_{\mathbf{k}},
\end{equation}
with
\begin{equation}
\gamma_{\rm eff} = 3H(\mathcal{L}_X + 3F + r),
\qquad
\omega_k^2 = (\mathcal{L}_X + 3F)\frac{k^2}{a^2}.
\end{equation}

This is structurally identical to standard warm inflation, with the replacement
\begin{equation}
\Gamma \rightarrow 3H(\mathcal{L}_X + 3F + r),
\end{equation}
showing that NMDC and noncanonical kinetic terms only rescale the effective dissipation.

\vspace{0.5em}

\subsection{Freeze-out Scale}

The relaxation time is
\begin{equation}
\tau_k = \frac{3H(\mathcal{L}_X + 3F + r)}{(\mathcal{L}_X + 3F)\,k^2/a^2}.
\end{equation}

Freeze-out condition $\tau_k \sim H^{-1}$ yields
\begin{equation}
k_F = H\sqrt{3(1+Q)},
\qquad
Q \equiv \frac{r}{\mathcal{L}_X + 3F}.
\end{equation}

This matches the standard warm inflation result up to the generalized dissipation ratio.

\vspace{0.5em}

\subsection{Thermal Variance from Stochastic Accumulation}

The inflaton variance is obtained from integration of modes up to freeze-out:
\begin{equation}
\langle \delta\phi^2 \rangle
= \int^{k_F} \frac{d^3k}{(2\pi)^3} \frac{T}{\omega_k^2}
\simeq \frac{k_F T}{2\pi^2}.
\end{equation}

This is identical in structure to the standard warm inflation result.

\vspace{0.5em}

\subsection{Comparison with Standard Warm Inflation}

\begin{center}
\begin{tabular}{c|c}
\hline
Standard Warm Inflation & NMDC Warm $k$-Inflation \\
\hline
$3H+\Gamma$ & $3H(\mathcal{L}_X + 3F + r)$ \\
$r=\Gamma/3H$ & $Q = r/(\mathcal{L}_X + 3F)$ \\
$k_F = H\sqrt{3(1+r)}$ & $k_F = H\sqrt{3(1+Q)}$ \\
$\delta\phi^2 \sim k_F T$ & identical form \\
\hline
\end{tabular}
\end{center}

This shows that the stochastic structure is unchanged; NMDC effects enter only through effective rescalings.

\vspace{0.5em}

\subsection{Consistency of Thermal Dominance}

The ratio
\begin{equation}
\frac{\delta\phi_{\rm th}^2}{\delta\phi_{\rm q}^2}
\sim \sqrt{3(1+Q)}\frac{T}{H}
\end{equation}
ensures that in the warm regime ($T>H$, $Q\gtrsim 1$), thermal fluctuations dominate, justifying the neglect of quantum contributions in the power spectrum calculation.

\vspace{0.5em}

\subsection{Conclusion}

The stochastic dynamics of the present NMDC warm $k$-inflation model are therefore a direct and controlled extension of standard warm inflation, with identical fluctuation structure that modified by effective friction rescaling. This fully justifies the analytic expressions used in Sec.~\ref{sec:level4}.


\end{document}